\documentstyle[12pt,epsf]{article}
\input axodraw.sty
\def\lsim{\:\raisebox{-0.5ex}{$\stackrel{\textstyle<}{\sim}$}\:}

\newcommand{\newc}{\newcommand}
\newc{\pbi}{pb$^{-1}$}
\newc{\ti}{\tilde}
\newc{\ra}{\rightarrow}
\newc{\ee}{$e^+e^-$\ }
\newc{\qq}{$q\bar{q}$\ }
\newc{\dd}{$d\bar{d}$\ }
\newc{\uu}{$u\bar{u}$\ }
\newc{\mm}{$\mu^+\mu^-$\ }
\newc{\taus}{$\tau^+\tau^-$\ }
\newc{\eeee}{$e^+e^-\ra e^+e^-$\ }
\newc{\eemm}{$e^+e^-\ra \mu^+\mu^-$\ }
\newc{\eett}{$e^+e^-\ra \tau^+\tau^-$\ }
\newc{\eeqq}{$e^+e^-\ra q\bar{q}$\ }
\newc{\eeuu}{$e^+e^-\ra u\bar{u}$\ }
\newc{\eedd}{$e^+e^-\ra d\bar{d}$\ }
\newc{\beq}{\begin{eqnarray}}
\newc{\eeq}{\end{eqnarray}}
\newc{\dqu}{\delta_{qu}}
\newc{\dqd}{\delta_{qd}}
\newc{\non}{\nonumber}
\newc{\noi}{\noindent}
\def\Rs{R \hspace{-0.38em}/\;}

\def\ib#1,#2,#3{       {\it ibid.\/ }{\bf #1} (19#2) #3}
\def\ap#1,#2,#3{       {\it Ann.~Phys.~(NY)\/ }{\bf #1} (19#2) #3}
\def\ijmp#1,#2,#3{     {\it Int.\ J.~Mod.\ Phys.\/ } {\bf A#1} (19#2) #3}
\def\mpla#1,#2,#3 {     {\it Mod.~Phys.~Lett.\/ } {\bf A#1} (19#2) #3}
\def\npb#1,#2,#3{       {\it Nucl.\ Phys.\/ }{\bf B#1} (19#2) #3}
\def\npps#1,#2,#3{     {\it Nucl.\ Phys.~B (Proc.~Suppl.)\/ }{\bf B#1}
                             (19#2) #3}
\def\plb#1,#2,#3{      {\it Phys.\ Lett.\/ }{\bf B#1} (19#2) #3}
\def\pr#1,#2,#3{       {\it Phys.\ Rev.\/ }{\bf #1} (19#2) #3}
\def\prd#1,#2,#3{      {\it Phys.\ Rev.\/ }{\bf D#1} (19#2) #3}
\def\prep#1,#2,#3{     {\it Phys.\ Rep.\/ }{\bf #1} (19#2) #3}
\def\prl#1,#2,#3{      {\it Phys.\ Rev.\ Lett.\/ }{\bf #1} (19#2) #3}
\def\pro#1,#2,#3{      {\it Prog.~Theor.\ Phys.\/ }{\bf #1} (19#2) #3}
\def\rmp#1,#2,#3{      {\it Rev.~Mod.~Phys.\/ }{\bf #1} (19#2) #3}
\def\sp#1,#2,#3{       {\it Sov.~Phys.~Usp.\/ }{\bf #1} (19#2) #3}
\def\zpc#1,#2,#3{      {\it Z.~Phys.\/ }{\bf C#1} (19#2) #3}
\def\appb#1,#2,#3{     {\it Acta Phys.\ Polon.\/ }{\bf B#1} (19#2) #3}

\begin{document}
\topskip 2cm 
\begin{titlepage}
\hfill IFT/19/98

\hfill hep-ph/9807312

\vspace{2cm}
\begin{center}
{\large\bf $R$-Parity Violating Signals at Existing
Colliders\footnote{Invited talk at the {\it 12th Les Rencontres de Physique
de la Vall\'ee d'Aoste}, La Thuile, Aosta Valley   March 1-7, 1998}} \\
\vspace{2.5cm}
{\large Jan Kalinowski} \\
\vspace{.5cm}
{\sl Institute of Theoretical Physics, Warsaw University\\
Ho\.za 69, Warsaw, Poland }\\
kalino@fuw.edu.pl
\vspace{2.5cm}
\vfil
\begin{abstract}
Formation of the $s$-channel slepton resonances at LEP2 or Tevatron at
current energies is an exciting possibility in $R$-parity violating
SUSY models.  Existing LEP2 and Tevatron data can be exploited to look
for sleptons, or to derive bounds on the Yukawa couplings of sleptons
to quark and lepton pairs.
\end{abstract}
\end{center}
\end{titlepage}

\section{Introduction}
Recently there was an increase of interest in the $R$-parity violating
supersymmetric model (RPV SUSY). It has been triggered at the
beginning of 1997 by observations at HERA of a number of events at
high $Q^2$, high $x$ in $e^+p$ scattering \cite{data96} above the
Standard Model (SM) expectations. Soon in a number of theoretical
papers the supersymmetry with broken $R$-parity has been put forward
as a possible explanation of these events \cite{squark}. It has been
speculated that the events are due to the $s$-channel squark
production. Although the great expectations of observing a genuine
signal of ``new physics'' have not been confirmed by the data
collected during the 1997 run of HERA \cite{data97}, experimental
situation still remains unsettled since the excess of ``anomalous
events'' is not yet washed out by the SM background.

The analyses of the RPV SUSY models in the light of HERA data have
reached interesting conclusions. First, they demonstrated that the
limits on combinations of RPV Yukawa couplings and masses of relevant
supersymmetric particles that have been derived from rare
processes\footnote{Note that these limits are derived with simplifying
assumptions that one (or at most two) RPV couplings are different from
zero at a time.} are very tight.  Second, that comparable limits for
some of the couplings/masses could be obtained directly from LEP
and/or Tevatron data to verify the theoretical attempts to explain
HERA data.  By now the results of LEP and Tevatron experiments
\cite{lepres,tevatron} put additional constraints for a consistent
squark interpretation of HERA events.

If squarks are too heavy to be produced at HERA, LEP or Tevatron,
great surprises nevertheless still might be ahead of us. Since in SUSY
GUT scenarios sleptons are generally expected to be lighter than
squarks, sleptons may show up at LEP2 and/or Tevatron even if squarks
are beyond the kinematical reach. Indirect effects due to
$t/u$-channel exchanges of sfermions in collisions of leptons and
hadrons might be observed although they are expected to be rather
small given the tight limits on the RPV couplings. Pair production of
sleptons via $R$-parity conserving mechanisms could also be closed
kinematically.  However, the direct formation of sfermion resonances
in the $s$-channel processes can produce remarkable events.  Sleptons
could be produced as $s$-channel resonances in lepton-lepton and
hadron-hadron collisions, and could decay to leptonic or hadronic
final states in addition to $R$-parity conserving modes. Therefore in
my talk I will concentrate on possible effects of $s$-channel slepton
resonance production on four-fermion processes in $e^+e^-$ collisions
\begin{eqnarray}
& & e^+e^- \rightarrow \tilde{\nu}\rightarrow \ell^+\ell^- \label{elel} \\ 
& & e^+e^- \rightarrow \tilde{\nu}\rightarrow q\bar{q} \label{ququ}
\end{eqnarray}
and in $p\bar{p}$ collisions 
\begin{eqnarray}
&& p\bar{p} \rightarrow \tilde{\nu}\rightarrow \ell^+\ell^-
\label{drel} \\
&& p\bar{p} \rightarrow \tilde{\ell}^+\rightarrow \ell^+\nu \label{lnu}  
\end{eqnarray}The results presented here
have been obtained in collaboration with H. Spiesberger, R.~R\"uckl
and P.~Zerwas \cite{snu,tev}.

\section{SUSY with $R$-parity violation}

The minimal $R$-parity conserving  supersymmetric extension (MSSM)
of the Standard Model is defined by the superpotential 
\begin{equation}
  W_R=Y_{ij}^e L_iH_1 E^c_j + Y_{ij}^d Q_i H_1D^c_j
  +Y_{ij}^uQ_iH_2U^c_j +  \mu H_1 H_2 \label{Rcons}
\end{equation}
where standard notation is used for the left-handed doublets of
leptons ($L_i$) and quarks ($Q_i$), the right-handed singlets of
charged leptons ($E_i$), up- ($U_i$) and down-type quarks ($D_i$), and
for the Higgs doublets which couple to the down ($H_1$) and up quarks
($H_2$); the indices $i,j$ denote the generations and a summation is
understood, $Y^f_{ij}$ are Yukawa couplings and $\mu$ is the Higgs
mixing mass parameter.

The superpotential $W_R$ respects a discrete multiplicative symmetry
under $R$-parity, which can be defined as \cite{FF}
\begin{equation}
             R_p=(-1)^{3B+L+2S}
\end{equation} 
where $B$, $L$ and $S$ denote the baryon and lepton number, and the
spin of the particle: all Higgs particles and SM fermions and bosons
have $R_p=+1$, and their superpartners have $R_p=-1$. The $R_p$
conservation implies that the interaction Lagrangian derived from
$W_R$ contains terms in which the supersymmetric partners appear only
in pairs. As a result, the lightest supersymmetric particle (LSP) is
stable and superpartners can be produced only in pairs in collisions
and decays of particles.

In the SM the $SU(2)\times U(1)$ gauge symmetry and Lorentz invariance
imply accidental $B$ and $L$ number conservation. Due to the larger
Lorentz structure, supersymmetric versions of the SM allow
renormalizable $B$ and $L$ violating operators involving scalars with
non-zero $B$ and $L$ charges. For example, the Higgs superfield $H_1$
can replace any of the $L_i$ in eq.~(\ref{Rcons}) since it has the
same quantum numbers as lepton superfields $L_i$. In general, the
gauge and Lorentz symmetries allow us to add the following terms to
the superpotential
\begin{equation}
  W_{\Rs}=\lambda_{ijk}L_iL_jE^c_k + \lambda'_{ijk}L_iQ_jD^c_k
  +\lambda''_{ijk} U^c_iD^c_jD^c_k\label{Rviol} +
   \epsilon_i L_i H_2 
\end{equation} 
which break explicitly the $R$-parity \cite{WSY}. If the Yukawa
couplings $\lambda$, $\lambda'$, $\lambda''$ and/or dimensionful mass
parameters $\epsilon$ are present, the model has distinct features:
superpartners can be produced singly and the LSP is not stable. Note
that at least two different generations of fermions are coupled in the
purely leptonic or purely hadronic operators.\footnote{Because of
anti-commutativity of the superfields, $\lambda_{ijk}$ can be chosen
to be non-vanishing only for $i < j$ and $\lambda''_{ijk}$ for $j<k$.
Therefore for three generations of fermions, $W_{\Rs}$ contains
additional 48 new parameters beyond those in eq.~(\ref{Rcons}).}

From the theoretical point of view, both types of
models, $R_p$-conserving or violating, have been constructed with no
preference for either of the two \cite{D1}.  Since they lead to very
different phenomenology, both models should be searched for
experimentally.  

The $\lambda$, $\lambda'$ and $\epsilon$ terms violate lepton number
and lepton flavor, whereas $\lambda''$ violate baryon number and
baryon flavor, and thus can possibly lead to fast proton decay if both
types of couplings are present. Therefore, additional symmetries are
required to enforce proton stability and to suppress $B$ and $L$
violating transitions. In the usual formulation of the MSSM they are
forbidden by $R_p$ and the proton is stable. However,
there is no theoretical motivation for imposing $R$-parity. Other
discrete symmetries can stabilize the proton without requiring the
$R_p$ to be conserved. For example, baryon-parity (defined as
$-1$ for quarks, and $+1$ for leptons and Higgs bosons) implies
$\lambda''=0$. In this case only lepton number (and lepton flavor) is
broken, which suffices to ensure proton stability. Lepton-number
violating operators can also provide new ways to generate neutrino
masses. Although in general the $\epsilon$ terms cannot be rotated away
\cite{drv}, here we will restrict the discussion to the MSSM 
with broken $R_p$ with the most
general trilinear terms in eq.~(\ref{Rviol}) that violate $L$ but
conserve $B$.

The Lagrangian for  $\lambda$ and $\lambda'$ parts of
the Yukawa interactions have the following form: 
\begin{eqnarray} {\cal
  L}_{\Rs}&=&\lambda_{ijk}\left[ \ti{\nu}^j_L\bar{e}^k_Re^i_L
+\overline{\ti{e}}^k_R (\bar{e}^i_L)^c\nu^j_L
+\ti{e}^i_L\bar{e}^k_R\nu^j_L\right. \non \\
&&\left. \mbox{~~~~~~~~~}
-\ti{\nu}^i_L\bar{e}^k_Re^j_L-\overline{\ti{e}}^k_R (\bar{e}^j_L)^c\nu^i_L
-\ti{e}^j_L\bar{e}^k_R\nu^i_L \right] + h.c.  \nonumber \\ 
&+&\lambda'_{ijk}\left[( \ti{u}^j_L\bar{d}^k_Re^i_L
+\overline{\ti{d}}^k_R(\bar{e}^i_L)^cu^j_L +\ti{e}^i_L\bar{d}^k_Ru^j_L
)\right. \nonumber \\ &&\left.
\mbox{~~~~~~~~~}-(\ti{\nu}^i_L\bar{d}^k_Rd^j_L+\ti{d}^j_L\bar{d}^k_R\nu^i_L
+\overline{\ti{d}}^k_R(\bar{\nu}^i_L)^cd^j_L) \right] + h.c.\mbox{~~~}
\label{effl}
\end{eqnarray} 
The notation is standard: $u_i$ and $d_i$ denote $u$- and
$d$-type quarks, $e_i$ and 
$\nu_i$  -- the charged leptons and neutrinos of the $i$-th
generation, respectively. The scalar partners are denoted by a
tilde and the superscript $c$ is for charge conjugated states. 
In the $\lambda'$ terms, the up (s)quarks in the $eud$ terms
and/or down (s)quarks in the $\nu dd$ may be Cabibbo rotated in the
mass-eigenstate basis.  As we will discuss mainly sneutrino induced
processes, we will assume the basis in which only the up sector is
mixed, $i.e.$ the $\nu dd$ is diagonal. 

If some of the $\lambda$ and $\lambda'$ couplings are non-zero, many
interesting processes might be expected at current and future
colliders in which this scenario could be explored. For example, the
$\lambda'_{ijk} E_i Q_j D_k^c$ operator could be responsible for the
$s$-channel production of squarks in $e^+p$ collisions at HERA (for
$i=1$), or sleptons in $p\bar{p}$ (for $j=k=1$ in valence quark)
collisions. The operator $\lambda_{1j1} L_1 L_j E^c_1$, on the other
hand, can lead to the $s$-channel sneutrino formation at LEP.

\section{Indirect limits on $\lambda$ and $\lambda'$ couplings}

At energies much lower than sparticle masses, $R$-parity breaking
interactions can be formulated as effective four-fermion contact terms
which in general mediate $L$-violating and FCNC processes. Since the
existing data are consistent with the SM, stringent constraints on the
Yukawa couplings and sparticle masses can be derived. If, however,
only some of the terms with a particular generation structure are
present in eq.\ (\ref{effl}), then the effective four-fermion
Lagrangian is not strongly constrained.  The couplings can also be
arranged in such a way that there are no other sources of FCNC than
CKM mixing in the quark sector.

\begin{figure}[htbp]
\begin{picture}(100,100)(-80,0)
\ArrowLine(10,60)(40,60)
\ArrowLine(40,60)(70,90)
\Photon(40,60)(60,40){2}{2}
\ArrowLine(60,40)(90,70)
\ArrowLine(90,10)(60,40)
\put(5,62){ $\tau^-$ }
\put(93,67){ $e^-$}
\put(51,51){ $W$}
\put(73,87){$\nu_\tau$}
\put(93,7){$\nu_e$}
\end{picture}
\begin{picture}(100,100)(-150,0)
%
\ArrowLine(10,60)(40,60)
\ArrowLine(70,90)(40,60)
\DashLine(40,60)(60,40){2}
\ArrowLine(60,40)(90,70)
\ArrowLine(60,40)(90,10)
\put(5,62){ $\tau^-$ }
\put(93,67){ $e^-$}
\put(51,51){ $\tilde{e}$}
\put(73,87){$\nu_e$}
\put(93,7){$\nu_\tau$}
\end{picture}
\caption{Tau decay via the SM $W$-boson exchange, and via the
\protect$\ti{e}$ due to the  \protect$L_1L_3E^c_1$ operator.} 
\label{taudec}
\end{figure}
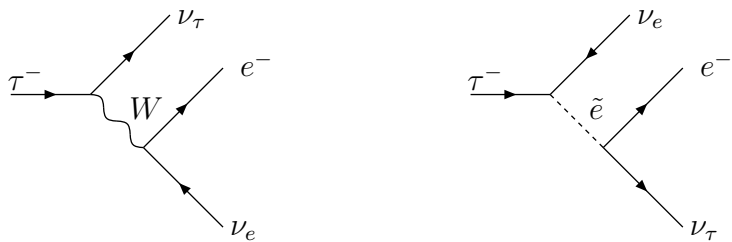

To illustrate how such constraints can be derived, let us consider a
specific example for which our group \cite{snu} contributed in
strengthening the experimental bounds denoted by $^b$ in Table~1. The
operator $\lambda_{131}L_1L_3E^c_1$ can contribute to the $\tau$
leptonic decay process $\tau\ra e\nu\bar{\nu}$ via the diagrams in
Fig.~\ref{taudec}.  After Fierz transformation the selectron exchange
diagram has the same structure as the SM $W$-boson exchange and thus
leads to an apparent shift in the Fermi constant for tau decays. The
ratio $R_\tau\equiv\Gamma(\tau\ra e\nu{\bar\nu})/\Gamma
(\tau\ra\mu\nu{\bar\nu})$ relative to the SM contribution is then
modified \cite{barger} 
\beq
R_\tau=R_\tau(SM)\left[1+2\frac{M_W^2}{g^2}
\left(\frac{|\lambda_{131}|^2}{{\tilde m}^2({\tilde
e}_R)}\right)\right]. \eeq 
Using the experimental value \cite{pdb} we obtain \cite{snu} the bound
\beq |\lambda_{131}| < 0.04\,
\left(\frac{{\tilde m}({\tilde e}_R)}{100~\mbox{GeV}}\right) \eeq

The Table 1 summarizes the strictest bounds on $\lambda$ and
$\lambda'$ couplings assuming that one RPV coupling at a time is
dominant while the others are neglected; bounds on products of two
couplings are not included. The bounds are given for sparticle masses
$\tilde{m}=100$ GeV. Those marked with $^*$ are based on a further
assumption about the absolute mixing in the quark sector. For more
details, discussion of physical processes from which they have been
obtained, and references we refer to \cite{D1, batta} from where
most of the entries of Table~1 have been taken.

\begin{table}[ht]
\begin{center}
\begin{tabular}{|cc||cc|cc|cc|}\hline
$ijk$ & $\lambda_{ijk}$ & $ijk$ & $\lambda'_{ijk}$ & $ijk$ & $\lambda'_{ijk}$ &
$ijk$ & $\lambda'_{ijk}$ \\ \hline
121 & 0.05     & 111 &0.00035   & 211 & ~~0.09~ & 311 & ~0.10~ \\
122 & 0.05     & 112 & 0.02     & 212 & 0.09 & 312 & 0.10  \\
123 & 0.05     & 113 & 0.02     & 213 & 0.09 & 313 & 0.10  \\
131 & 0.04$^b$ & 121 & 0.035    & 221 & 0.18 & 321 & 0.20$^*$ \\
132 & 0.04$^b$ & 122 & 0.02     & 222 & 0.18 & 322 & 0.20$^*$  \\
133 & 0.004    & 123 & 0.20$^*$ & 223 & 0.18 & 323 & 0.20$^*$  \\
231 & 0.04$^b$ & 131 & 0.035    & 231 & 0.22 & 331 & 0.26  \\
232 & 0.04$^b$ & 132 & 0.33     & 232 & 0.39 & 332 & 0.26 \\
233 & 0.04$^b$ & 133 & 0.001    & 233 & 0.39 & 333 & 0.26 \\
\hline
\end{tabular}
\end{center}
\caption{Bounds on RPV Yukawa couplings for ${\tilde m}=100$ GeV.
\label{tab:bounds}} 
\end{table}

In the discussions to follow we will consider two specific scenarios:
\\[1mm] ($i$) one single Yukawa coupling is large, all the other
couplings are small and thus neglected; \\[1mm] ($ii$) two Yukawa
couplings which violate {\it one and the same} lepton flavor are
large, all the others are neglected.

First we shortly recapitulate the situation concerning the HERA data,
$i.e.$ the effects that can be generated by $L_1 Q_j D_k^c$ operator.
Then we will discuss slepton production at LEP and Tevatron. In this
context we will concentrate on possible effects generated by
$\tilde{\tau}$ and $\tilde{\nu}_{\tau}$, ($i.e.$ $\lambda_{i3i}$ and
$\lambda'_{3jk}$ couplings) since the third-generation sfermions are
usually expected to be the lightest and, due to large top quark mass,
the violation of the third-generation lepton-flavor might be expected
maximal. In these cases low-energy experiments are not very
restrictive, see Table~1, and typically allow couplings to be of the
order 0.1 for the mass scale 200 GeV of the sparticles participating
in the process.

\section{$L_1 Q_j D_k^c$ operator}

In $e^+p$ collisions at HERA the operator $L_1 Q_j D_k^c$ could be
responsible for squark resonance production via
\begin{equation}
e^+ d_R^k \rightarrow \tilde{u}_L^j
~~~~~~ (\tilde{u}^j = \tilde{u}, \tilde{c}, \tilde{t}),
\label{kwark}
\end{equation}
\begin{equation}
e^+ \bar{u}_L^j \rightarrow \overline{\tilde{d}}_R^k
~~~~~~ (\tilde{d}^k = \tilde{d}, \tilde{s}, \tilde{b}).
\label{antykwark}
\end{equation}
Given the bounds in Table~1, charm or top squarks can be produced off
the $d$ quarks via eq.~(\ref{kwark}). Since the excess of events was
only observed in $e^+p$, not in $e^-p$ scattering, the process induced
by $\bar u$ sea in eq.~(\ref{antykwark}) is unlikely. For the
production off other sea quarks, where the coupling strength $ \simeq
e$ is required, only stop production off strange sea is still compatible
with the existing bounds.  In short, three possible explanations of
the HERA anomaly have been identified \cite{squark}
\begin{eqnarray}
e^+d \to \tilde{c} ~~(\lambda'_{121}), \label{dscharm} \\
e^+d \to \tilde{t} ~~(\lambda'_{131}), \label{dstop}  \\
e^+s \to \tilde{t} ~~(\lambda'_{132}). \label{sstop}
\end{eqnarray}
Within the limits on $\lambda'$ in Table~1, branching ratios $B_{eq}$ for
$\tilde{c}, \tilde{t} \rightarrow e^+d$ should fall below 0.7 in order
to avoid the D0/CDF mass bounds \cite{tevatron}. It has been shown 
in \cite{sqtev}
that one can indeed find solutions in the supersymmetry parameter
space in which $B_{eq} < 0.7$, although the allowed region for a
consistent squark interpretation of the HERA anomaly and LEP and D0/CDF 
bounds is
very limited. RPV SUSY may also provide a reasonable solution
\cite{kon} of the difficulty to interpret the excess of events as a
single-resonance effect: mixing in the stop sector may lead to two
mass eigenstates with a small but pronounced mass difference,
mimicking a continuum effect.

The NC events from $\tilde{t}, \tilde{c} \to e^+ d$ have the same
visible final states as the standard DIS-NC events.  This is not the
case for CC events since the left squarks produced in processes
(\ref{dscharm}--\ref{sstop}) do not couple to neutrinos and quarks,
see eq.~(\ref{effl}).  CC-like events could only originate from
cascade decays of squarks with some jets in the final state either
invisible or overlapping. The H1 events with isolated muons and
missing transverse momentum \cite{muon} are difficult to explain.

The $L_1 Q_j D_k^c$ operator could also contribute to processes at LEP
via $t$- or $u$-channel exchange of sparticles, although the effects
are expected to be small \cite{lq} for the couplings listed in
Table~1. In contrast, in $p\bar{p}$ collisions sleptons can be
produced in the $s$ channel via the $LQD^c$ operator with appreciable
cross section.  In hadronic environment, however, the decay modes
induced by either $R_p$-conserving gauge or $R_p$-violating Yukawa
$\lambda'$ couplings might be quite difficult. On the other hand, if
$LLE^c$ operators are present, the leptonic decay modes can be easily
detected, as discussed in the next section.

\section{$L_i L_j E_k^c$ operator}
 
In \ee scattering at LEP sleptons can be produced singly in the
$s$-channel via $LLE^c$ and in $p\bar{p}$ at Tevatron via $LQD^c$
operators leading to a number of different signatures depending on the
assumed scenario.\footnote{Sleptons can be also exchanged in the $t$
or $u$ channels, see below.}  Once produced, they can decay via either
the $R_p$-violating Yukawa or the $R_p$-conserving gauge couplings. In
the latter case the decay proceeds in a cascade process which involves
standard and supersymmetric particles in the intermediate states and
with the $R_p$-violating coupling appearing at the end of the
cascade. Such decay processes lead in general to multibody final
states and depend on many unknown SUSY parameters.  In the former
case, the final state is a two-body state (with two visible particles,
eqs.~(\ref{elel})-(\ref{drel}), or one visible particle and a missing
momentum, eq.~(\ref{lnu})) which depends only on a limited number of
parameters and which is very easy to analyze experimentally. Therefore
we will consider sleptons that are produced and decay via $\lambda$
and/or $\lambda'$ couplings, namely their effects on four-fermion
processes at LEP and Tevatron.

The general expressions for a generic two-body
process\footnote{$f,\bar{f}',F$ and 
$\bar{F}'$ are SM fermions.}  $f\bar{f}'\rightarrow
F\bar{F}'$, including the exchanges of sparticles in $s$, $t$ and/or
$u$ channels, can be found in \cite{tev,ustron}. For the energy close
to the mass of the sparticle $\ti{p}$ exchanged in the $s$ channel,
the cross section is well approximated by the Breit-Wigner formula
\begin{equation}
\sigma(f\bar{f}'\rightarrow \ti{p} \rightarrow F\bar{F}')= 
\frac{4\pi s}{m^2_{\ti{p}}}\;
\frac {\Gamma(\ti{p}\ra f\bar{f}' )\Gamma(\ti{p}\ra F\bar{F}')}
{(s-m^2_{\ti{p}})^2+m_{\ti{p}}^2\Gamma^2_{\ti{p}}} 
\end{equation}

\begin{figure}[htbp] 
\unitlength 1mm
\begin{picture}(90,90)
\put(0,-100){
\epsfxsize=15cm
\epsfysize=18cm
\epsfbox{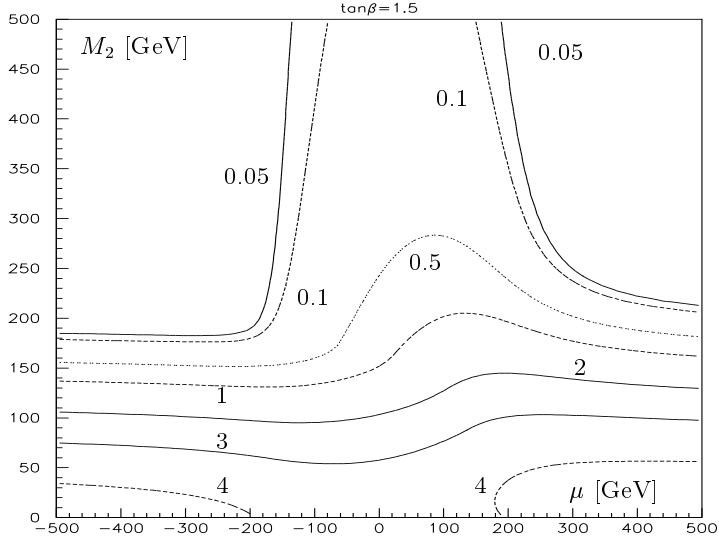}}
\end{picture}
\caption{ Contour lines for the sneutrino total decay width (in GeV)
   as a function of gaugino mass $M_2$ (gaugino unification assumed)
   and Higgs mixing parameter $\mu$. The sneutrino mass
   $m_{\ti{\nu}}=200$ GeV and $R_p$ violating couplings
   $\lambda=\lambda'=0.08$ are assumed, and $\tan{\beta}=1.5$.  }
\label{width}
\end{figure}

The partial width for $R_p$-violating decay 
$\Gamma(\ti{p}\ra f\bar{f'})=\lambda^2
m_{\ti{p}} /16\pi$ is very small for Yukawa 
couplings consistent with Table~1.  However, the total decay 
width $\Gamma_{\ti{p}}$ can be much larger since sparticles can
also decay via $R$-parity conserving gauge couplings. As an example we
will consider sneutrinos.  They can decay to $\nu\chi^0$ and
$l^{\pm}\chi^{\mp}$ pairs with subsequent $\chi^0$ and $\chi^{\pm}$
decays and via $R$-parity violating $\lambda'$ couplings to
$q\bar{q}$, or via $\lambda$ couplings to lepton pairs.  The partial decay
widths for these channels depend on the specific choice of the
supersymmetry breaking parameters.  In large regions of the
supersymmetry parameter space, the total decay width of sneutrinos can
be as large as 1 GeV, as illustrated in Fig.~\ref{width}. It means
that at LEP2 the sneutrino total decay width can be significantly
larger than the beam energy spread.  Therefore the interference
with the background Standard Model process must be taken into account.

\subsection{Sneutrinos in \ee Scattering }
As discussed earlier we consider the couplings that violate 3rd
generation flavor number, $\lambda_{131}$, $\lambda_{232}$ and
$\lambda'_{3jk}$. Several processes can be affected in such a
scenario:\\[1mm]
\noindent (a) {\it Bhabha scattering:} For $\lambda_{131}\ne 0$,  
the tau sneutrino $\ti{\nu}_{\tau}$ can
contribute to Bhabha scattering via $s$- and $t$-channel
exchanges.  Note that the $s$-channel ($t$-) sneutrino
exchange interferes with the $t$-channel ($s$-) $\gamma,Z$
exchanges. 
\\[1mm]
\noindent (b) {\it Muon-pair production:} This process can be mediated
by the $s$-channel $\ti{\nu}_{\tau}$ resonance, $e^+e^-\ra\mu^+\mu^-$,
if in addition $\lambda_{232}\ne0$.  Since the $t$-channel $\gamma,Z$
and $\ti{\nu}_{\tau}$ exchanges are absent, the $s$-channel sneutrino
exchange does not interfere with the SM processes.\\[1mm]
\noindent (c) {\it Tau-pair production:} 
This process can receive only
the $t$-channel exchange of $\ti{\nu}_e$ 
which will interfere with the SM $\gamma,Z$ $s$-channel
processes.\\[1mm] 
\noindent (d) {\it Neutrino-pair production:} Electron (tau) neutrinos
can receive additional contributions only via $t/u$-channel exchanges
of $\ti{\tau}$ ($\ti{e}$), which will interfere with the SM $Z$-exchange 
process.  \\[1mm]
\noindent (e) {\it $e^+e^-$ annihilation to hadrons:}
The up-type quark-pair production is not affected by sneutrino
processes, as can be easily seen from the general structure of
couplings in eq.~(\ref{effl}). For the
down-type quark-pair\footnote{The possibility of $\tilde{\nu}_\tau
\rightarrow b\bar{b}$ has been discussed in the context of $e^+e^-
\rightarrow b\bar{b}$ at LEP1 \cite{fnp}.}
 production, $e^+e^-\ra d_k\bar{d}_k$, 
the situation
is similar to the muon-pair production
process: there is no interference between $s$-channel
$\ti{\nu}_{\tau}$ exchange and the SM $\gamma,Z$ processes. 
The unequal-flavor
down-type quark-pair production process, $e^+e^-\ra
d_j\bar{d}_k$, could be generated only by $s$-channel sneutrino with 
$\lambda_{131}\lambda'_{3jk}\ne 0$.

\begin{figure}[ht] 
\unitlength 0.8mm
\begin{picture}(100,110)
  \put(10,-70){ \epsfxsize=14cm \epsfysize=15cm\epsfbox{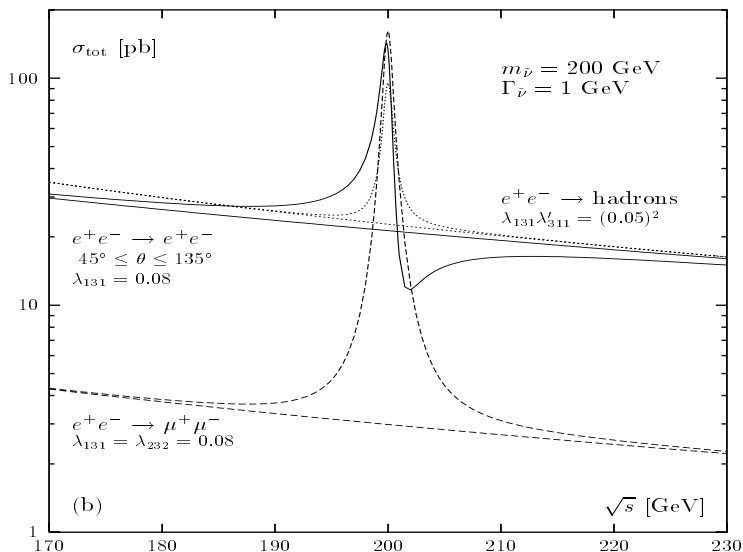}}
\end{picture}
\caption{ Cross section for Bhabha scattering
  (solid lines), $\mu^+\mu^-$ (dashed lines) and hadron production
  (dotted lines) in the SM, and including $\ti{\nu}_{\tau}$ sneutrino
  resonance formation as a function of the $e^+e^-$ energy. }
\label{figlep}
\end{figure}
In general, the effect of $t$- or $u$-channel exchange of sleptons is
very small (typically below 1\%) for the slepton masses and couplings
consistent with low-energy data. On the other hand, in processes with
$s$-channel exchanges, and not too far from the resonance, the effect
of sneutrino can be quite spectacular. This is illustrated in
Fig.~\ref{figlep}, where the impact of the exchange of tau sneutrino
with $m_{\ti{\nu}_{\tau}}=200$ GeV and $\Gamma_{\ti{\nu}_{\tau}}=1$ GeV on
processes (a), (b) and (e) at LEP2 is shown. Note the
difference due to different interference pattern between Bhabha
scattering on one hand, and muon-pair and quark-pair production
processes on the other: Bhabha is more sensitive to heavy sneutrinos.
The peak cross section for Bhabha scattering is given by the unitarity
limit $\sigma_{peak}=8\pi B_e^2/m^2_{\ti{\nu}_{\tau}}$ with sneutrino
and anti-sneutrino production added up, where $B_e$ is the branching
ration for the sneutrino decay to $e^+e^-$.  The cross section in the
peak region is therefore very large. Another important feature of the
sneutrino resonance is the change in the angular distribution of
leptons and quark jets: the distribution is nearly isotropic with the
strong forward-backward asymmetry in the Standard Model continuum
reduced to $\sim 0.03$.  In addition to $\ell^+ \ell^-$ and
$q\bar{q}$ final states one should expect many other final states
generated in $R$-parity conserving $\ti{\nu}$ decays to $\nu\chi^0$
and $\ell^\pm\chi^\mp$ pairs with subsequent $\chi^0$ and $\chi^\pm$
decays \cite{godbole}.

\begin{figure}[htbp] 
\unitlength 0.8mm
\begin{picture}(100,120)
  \put(10,-10){ \epsfxsize=12cm \epsfysize=12cm\epsfbox{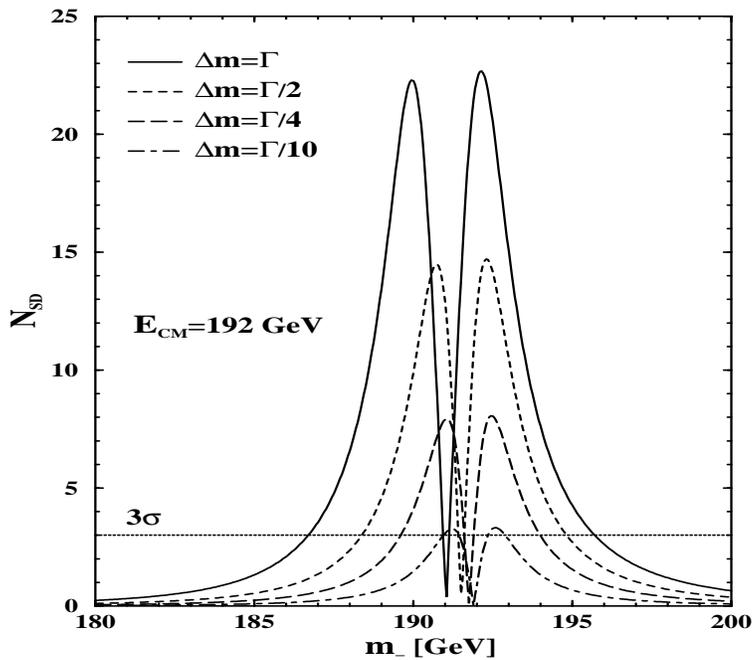}}
\end{picture}
\caption{ The statistical significance, $N_{SD}$, attainable at LEP2
for spin asymmetries $A_{xy}$ and $B$ as a function of the lighter
muon sneutrino mass for several values of the mass splitting. Figure
taken from Ref.~[22] to which we refer for details.}
\label{spinas}
\end{figure}

An interesting situation may occur if sneutrinos mix and mass
eigenstates are split by a few GeV \cite{snumix}.  Then one may
expect two separated peaks with reduced maximum
cross sections 
in the energy dependence in Fig.~\ref{figlep} for 
processes (a), (b) and/or (d). If the mass splitting is below
the energy resolution, 
one may nevertheless resolve sneutrino mass eigenstates by measuring
CP-even and CP-odd spin asymmetry of final states leptons
\cite{eilam}. From the experimental point of view such measurements
can be done only for $\tau$ pairs using spin self-analyzing decay
modes.  In the scenarios considered so far, however, $\tau$-pair
production is not affected by $s$-channel $\ti{\nu}_{\tau}$
process. If instead of $\tau$-flavor the muon-flavor is violated via
$\lambda_{121}\lambda_{233}\ne 0$, then the asymmetries in $e^+e^-
\rightarrow \ti{\nu}_{\mu} \rightarrow \tau^+\tau^-$ can be measured
with high statistical significance, as shown in Fig.~\ref{spinas}
taken from Ref.~\cite{eilam}.

\subsection{Sleptons at Tevatron}
For $p\bar{p}$ scattering 
the case $\lambda'_{311}$ is the most
interesting since it allows $\ti{\nu}_{\tau}$ and $\ti{\tau}$ 
resonance formation in valence quark collisions. 
As their decays to quark jets can be very difficult to observe in
hadronic environment,  
we will consider leptonic decays of sleptons via
$\lambda_{i3i}$ couplings. To be specific, we take
$\lambda_{131}$ and discuss $e^+e^-$ and $e^+\nu_e$ production
in $p\bar{p}$ collisions; the same results hold for $\mu^+\mu^-$
and $\mu^+\nu_{\mu}$ production if $\lambda_{232}$ is
assumed. The differential cross
sections for $p\bar{p}\ra e^+e^-$ and $e^+\nu_e$ processes 
are obtained by combining the luminosity spectra for
quark-antiquark annihilation with partonic cross sections for\\[1mm]
\noindent (a) {\it electron-pair production:} the $s$-channel
sneutrino $\ti{\nu}_{\tau}$  
exchange contributes
only to $d\bar{d}\ra e^+e^-$   
which does not interfere with  the SM $s$-channel $\gamma,Z$
processes; \\[1mm]
\noindent (b) {\it electron + missing energy:} only the
process $u\bar{d}\ra e^+\nu_e$   
receives the 
$s$-channel $\ti{\tau}$ slepton exchange which does not
interfere with  the $s$-channel $W$-boson exchange. 

\begin{figure}[htbp] 
\unitlength 0.8mm
\begin{picture}(100,120)
  \put(10,-55){ \epsfxsize=14cm \epsfysize=16cm \epsfbox{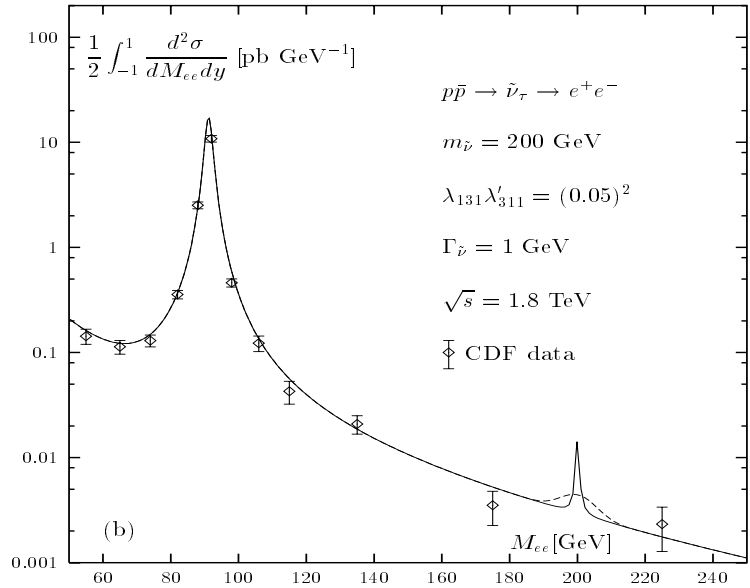}}
\end{picture}
\caption{The $e^+e^-$ invariant mass distribution including the $s$-channel
  sneutrino in the channel $d\bar{d}\ra e^+e^-$ is compared with the
  CDF data; solid line: ideal detector, dashed line: sneutrino
  resonance smeared by a Gaussian width 5 GeV. The CTEQ3L structure
  functions have been used. }
\label{figtev}
\end{figure}

In numerical calculations the total decay widths
of sleptons have been set to a typical value of 1 GeV, corresponding
to the branching ratios for leptonic decays of order 1\%. The
resulting 
di-electron invariant mass distribution is compared to the CDF data in
Fig.~\ref{figtev}. Following CDF procedure \cite{CDF},
the prediction for 
$\frac{1}{2}\int^{1}_{-1}
{\mbox{d}^2\sigma}/{\mbox{d}M_{ee}\mbox{d}y}$ is shown. The
solid line is for an ideal detector, while the dashed line 
is for the distribution after the smearing of the peak by
experimental resolution characterized by a Gaussian width of 5 GeV. 
The CTEQ3L parametrization \cite{cteq} is used together
with a multiplicative $K$ factor for higher order QCD corrections to
the SM Drell-Yan pair production.

\section{Summary}

The $R$-parity violating formulation of supersymmetric extension
of the SM  offers a distinct
phenomenology and therefore deserves detailed studies. Even if the
squarks are beyond the kinematical reach of HERA, sleptons might be
light enough to be produced as $s$-channel resonances with
spectacular signatures at LEP2 and/or Tevatron. We
discussed the scenario with lepton number violation,  and we 
enumerated a number of processes in which sleptons might play an important
role. We concentrated only on four-fermion processes in which sleptons are
produced and decay via $R_p$-violating couplings.  On the other hand, if no
deviations from the SM expectations are observed, stringent bounds on
individual couplings can be derived experimentally in a direct way.
 For example,
if the total cross section for $e^+e^-$ annihilation to hadrons
at 192 GeV can be measured to an 
accuracy of 1\%, the Yukawa couplings for a 200 GeV sneutrino can be
bounded to $\lambda_{131}\lambda'_{311}\lsim (0.045)^2$ \cite{tev}.
Similarly, assuming the sneutrino contribution to di-electron
production at Tevatron be smaller than the
experimental errors, we estimate that the bound
$\lambda_{131}\lambda'_{311}\lsim (0.08)^2 \ti{\Gamma}^{1/2}$ can be
established \cite{tev}, where $\ti{\Gamma}$ denotes the sneutrino width in
units of GeV. \\[2mm]

\noindent {\bf{\large Acknowledgments}}\\[2mm]
I would like to thank Mario Greco and Giorgio Bellettini 
for the invitation to La Thuile and warm hospitality at the conference.
I am grateful to R.~R\"uckl, H.~Spiesberger and P.~Zerwas
for enjoyable collaboration and many discussions and comments. 
I also thank    
W.~Krasny,  Y.~Sirois and F.~\.Zarnecki for
discussions and communications. The work has been partially supported 
by the Polish Committee for Scientific Research Grant 2 P03B 030 14.

\end{document}